\documentclass[12pt]{article}
\usepackage{pdproc}
\usepackage{pdproc,epsfig}
\begin{document}

\renewcommand{\thefootnote}{\alph{footnote}}

\title{
ACHIEVEMENTS AND MIRAGES in UHECR and NEUTRINO  ASTRONOMY }

\author{ D.Fargion, D. D'Armiento}

\address{ Physics Department and INFN, Rome University 1, Sapienza,
  Address\\
 Ple A. Moro 2, 00185, Rome. Italy\\
 {\rm E-mail: daniele.fargion@roma1.infn.it}}
  \centerline{\footnotesize }
\abstract{  Photon Astronomy ruled the last four centuries while wider photon band ruled last century of discovery. Present decade may see the rise and competition of UHECR and UHE Neutrino Astronomy. Tau Neutrino may win and be the first flavor revealed. It could soon rise at horizons in AUGER at EeV energies, \emph{if nucleons are the main UHECR  currier}. If on the contrary UHECR are Lightest nuclei (He, Li. B)  UHE tau neutrino maybe suppressed at EeV and enhanced at tens -hundred PeV. Detectable in AMIGA and HEAT denser sub-array in AUGER. Within a few years.}

\normalsize\baselineskip=15pt
\section{Introduction: UHECR versus UHE neutrino Astronomy}

\begin{figure}
\centering
\includegraphics[width=0.9\textwidth]{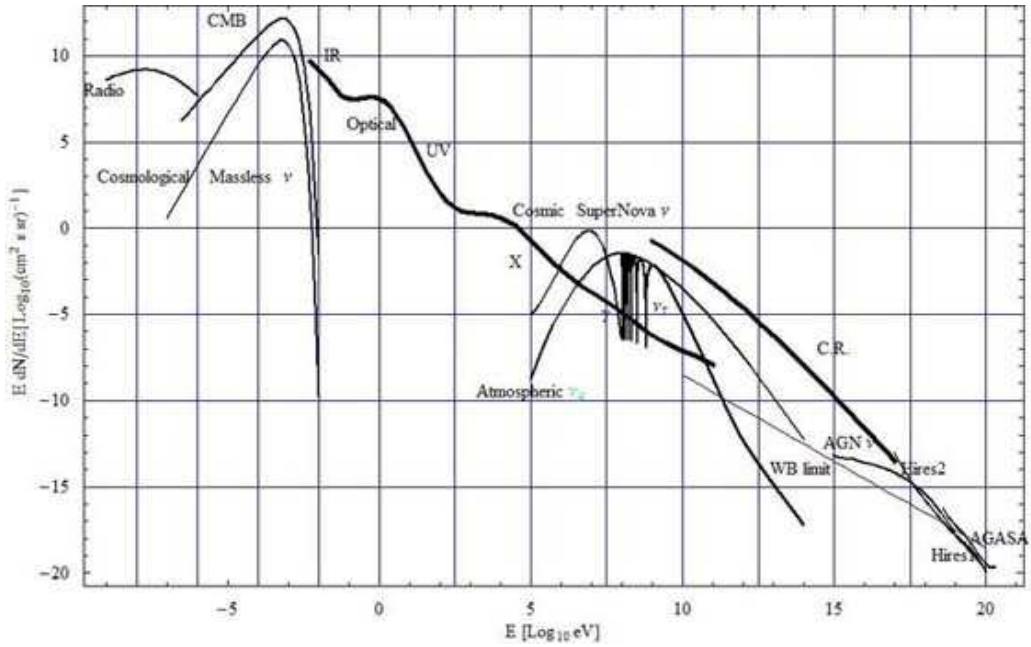}
\caption{The Cosmic Radiations: the lower energy left side spectra mostly describe photons. Higher energy (above TeVs) radiations on right side
are describing the charged Cosmic ray contribute. At these high energies (TeVs-PeVs) photons are scattering onto Infra-red, optics and BBR ones, making opaque far cosmic sources. Similar energetic neutrinos are transparent to such photons. However at Earth both electron and muon neutrinos suffer of the secondary atmospheric noise. Tau are well produced by oscillation (shown above in logarithmic fashion). The higher the energy, the less the oscillation. At  TeV energies tau neutrinos are almost absent because the small Earth size respect to the neutrino-mass oscillation length. This make tau neutrino noise free, but very short and hard to observe, at TeVs. However  at PeVs-EeVs energies the tau track are more and more long and rising deeper from matter below. Their later decay in flight is better amplified and disentangled from any (absent) hadron air-shower at horizons. An inclined line marked Waxmann Bachall (WB) label the minimal energy (at Tens EeV) fluency in CR. Incidentally comparable to GRB main  fluency. This is the minimal fluency of PeVs-EeVs neutrinos in the Sky.If  UHECR are not nucleons EeV neutrinos may be less present,their flux may escape as a mirage. See Last Figures}\label{fig:fig01}
\end{figure}

Galileo four century ago invented large lens  telescopes for an amplified view of the stars. Its  strategy ruled and rules optical Astronomy. More photon telescopes in wider spectra, from largest radio antenna to space X-gamma satellite, extended our views last century in Maxwell regimes. However at hundred TeVs photons hit abundant relic cosmic photons making electron pairs. This opacity leads to a very bounded  tens TeVs gamma   Astronomy almost galactic at PeV energy  band. Cosmic Rays offer an energetically flux comparable to photon ones but they are charged and bent by cosmic magnetic fields. Therefore  CR astrophysics is blurred and blind.   UHECR are as much as energetic to offer a straight view of the source and a novel Astronomy. But also UHECR at GZK cut off (tens EeV energy) are bounded in a very Local Universe too. As small as $1\%$ radius and a part of a million of the volume. First spectra and  maps of AUGER UHECR seemed to see the first Super-Galactic shadows, or maybe as we critically noted, just their mirages formed by the main Cen-A spread events,  as a main blurred source in  AUGER map. Virgo absence is obvious by He-Li-Be opacity beyond $10$ Mpc. The same UHECR GZK  secondary, the cosmogenic neutrinos are neutral and are tracing, with no photon-like opacity, wider  cosmic spaces. Better and deeper tracing explosive or Jet source. This push most to call for  UHE Neutrino Astronomy. Electron neutrino are ruling only at few-tens-hundred MeV.  GeVs muons are more penetrating and better detectable. The muon  neutrino search above tens-hundreds GeV, implemented in km cube ice or water underground, is already  four decade old. Its records are slowly growing now, finally reaching  this year its dreamt performance. This is a great achievements of the decade. But muon neutrinos are polluted vertically  by a noisy atmospheric neutrino background. At less polluted higher energy UHE $\nu_{\mu}$ are suppressed by Earth Opacity. At horizons, while the Earth cord is shorter and less opaque to UHE neutrinos, the prompt atmospheric neutrinos $\nu_{\mu}$, $\nu_e$ are more abundant, by at least an order of magnitude, than vertical ones. Making a even more noisy the sky to $\nu_{\mu}$, $\nu_e$ . On the other side the tau Neutrino Astronomy  invented only a decade ago, is much more silent and quite. Indeed it is already competing with the muon one via AUGER bounds, reaching AMANDA and maybe ICECUBE. Tau are extremely difficult to be produced.  But the muon neutrino flavor mixing  induce and rise  their rare flavor. But the oscillation do not occur in Earth size above tens GeV.  The $\nu_{\tau}$ are the neutral lepton of the heaviest and most unstable charged $\tau$. Such a Ultra High Energy (UHE) tau born after neutrino skimming and interacting on the Earth skin may lead, once in air,  to Tau Air-shower. A huge amplified explosion in air. Spraying over huge area huge number of secondary particles.  Possibly the first Neutrino Astronomy to be discovered. Because of the severe Earth opacity, EeV neutrinos grow at best at horizons (short cord) \emph{below} the horizons. There are no rich  hadrons  air-shower filtered at horizons or even below.  The $\nu_{\tau}$ sky lay beyond the mountains chains or under our feet \cite{Fargion1999}. Mostly in AUGER area, as the wide nearby Ande screen.  Energetic  Tau neutrinos above TeV energy are  noise free respect overabundant  atmospheric  muon neutrino. But taus at TeVs are short in matter  and rare to emerge. Their  skimming Earth at PeVs-EeVs energy  and the consequent tau decay in air offer however an unique neutrino air-showering upward-horizontal, at low atmosphere  ($1-8$ km) , in a few km deep and few microsecond long track; they differ drastically from hadronic tangent air-showers blowing only at very high (25-35)  km diluted altitude. Such hadron air-shower are very thin, lengthy  (hundreds km) and longevous  showers. Difficult (or impossible) to observe in AUGER Fluorescence Detector. The hadron air-shower beams are often split by geomagnetic fields.  Therefore the Fluorescence detection\emph{ of upward EeV Tau} air-shower is at hand (this or next years) in largest fluorescence telescope array as AUGER,  \emph{if UHECR are mostly nucleons}. However  UHECR maps and their  observed composition puzzle us. They  forced us to imagine  \emph{UHECR as the lightest nuclei}: their consequent UHECR  GZK secondary neutrinos, are only partially born  (half a number) in photo-pion. Their secondary neutrinos must be mostly  by  nuclear fragmentation within a  very Local Universe (Cen-A being the main source  at $4$ Mpc). These UHECR processes   are making fewer GZK neutrinos than expected at EeV, reducing as in a mirage their detection  goal. More abundant signals at tens-hundred  PeV energy must nevertheless   rise. These UHE PeVs tau neutrinos might be  revealed better in a new inner and denser  array of AUGER,  the HEAT and AMIGA, tuned to lower energies. Or in competition within new born ICE-CUBE km  detector mostly via UHE lengthy muons (if disentangled by UHE prompt neutrinos). Double bang may also rise.  We foresee them to be found soon, within a few  years (3-4), in a very peculiar and marked imprint in AUGER-HEAT-AMIGA array. Possibly tracing the resonant  Glashow anti-neutrino electron.

\section{ New Neutral Particle Telescopes?}
 Look at the blue sky in day time: no star are visible in diffused sun lights. To see stars one must look at dark nights. This remind us that  a signal shine at best in  a noise free screen. To better reveal any weak signal also amplifiers are needed. Optical photons (mostly linked to sun spectra) forced astronomers on top altitude observatories. Galileo used the lens not just for a  sharper view but also for an amplified one. Photons are neutral and they flight along a direct line. Offering a clear view. Cosmic Rays, as rich as other cosmic fluency  are charged ones. They suffer of bending and blurring along the cosmic (Earth, Sun, Galactic, intergalactic) magnetic fields. The same existence of large magnetic fields stand and remind us a neglected cosmic puzzle: the (apparent) absence of any magnetic monopole as cosmic relic trace. Photons are not the unique neutral particles: also neutrons and antineutrons and gravitons  are well known neutral particles. Indeed Solar flare (Compton Observatory,OSSE,1991)  shine GeVs neutrons at largest solar flares. PeV Neutrons may also rise from ten parsec nearest pulsars. EeV neutrons may survive even from our galactic center. A decade ago such an anisotropy was revealed and  claimed from AGASA, but it has been disproved by later AUGER records. Tens-hundred EeV neutrons may rise by GZK secondaries even from nearest extragalactic sources (Mpc distance). Antineutrons are much rarer: possibly born in a very speculative anti-galaxy  via anti GZK secondaries. Finally gravitons may also tremble in detectors as soon as  a Supernova shine in nearby universe. But at the present with little angular resolution. Finally speculative UHE neutralinos might shine if SUSY occurs and if it plays a comparable role in UHECR components. Apart these four neutral candidates, there are other guaranteed ones to be revealed. Neutrino Astronomy  is not just  one, but as their lepton flavors they  are three or better because anti-matter states, the neutrino astronomy are six.
\section{Six Neutrinos Astronomy: Virtues and Defects}

 Electron  are the lightest leptons and the easiest to observe because of the low  mass threshold. Inverse beta decay is a key processes. The nuclear plant were their first sources laboratory.  The shining overabundant Solar neutrino astronomy has been revealed last  decades via different detectors because their huge fluency. Supernova 1987A, even from Magellanic Cloud outside our Milky-Way,  had left a tiny signals in earliest detectors more than two decade ago. Cosmic Relic supernova (average traces) are at the edge of detections. A new and exciting neutrino astronomy might rise via Megaton underground detectors by largest solar flare.
 However at tens MeVs and above the astrophysical neutrinos are sinking into the atmospheric secondary sea.   Their corresponding Telescopes usually  lay in underground to screen the huge polluting downward cosmic ray lepton secondaries. Their direct lepton downward noise is nearly twelve order above any (tens GeV) neutrino signal. Moreover at high energy  electron and positron signals maybe overcome by muon ones because at GeVs energies and above muons are more penetrable and much longer than electrons. However muon (as well electron) neutrinos suffer of the ruling CR secondary noise. The CR secondary muons pollute abundantly the down ward vertical direction, even in deep km underground detectors. Nevertheless also
  Up-going neutrinos are polluted by the same atmospheric ones at least by three order of magnitude above  expected astrophysical sources.  At tens TeV energies the astrophysical signal may finally compete with the atmospheric noise. But Earth opacity rise at vertical axis. Tau are almost absent because charmed mesons are hard to be born.  However at GeVs muon neutrino oscillate into tau ones; but above ten GeVs their energies are so high that the oscillation cannot take place inside our Earth diameter. Therefore Tau neutrinos above TeV are mostly  of astrophysical nature with little or none mixing, just noise free. This make them interesting. As energy increase above PeVs tau length grows linearly while muon track grows in logarithmical law. Therefore at EeV Tau may be longer than muon track.  Moreover just above PeV energy (corresponding $50$ m tau distances) their birth first and they sudden decay after into Tau offer a very rare signature: a double bang ( Learned and Pakwasa (1995)) imprint in water or ice.  Therefore their relevance survive and rise above PeV.  At PeV-EeV energies their bang (in the Earth) and their decay out, (in the air) offer the Tau skimming decaying  in flight, producing loud, amplified  Tau air-showers. The muons searched in Cubic km detector  may be at detection edges if the UHECR are of lightest nuclei nature.   However as noted above of  neutrino pollution ,  at horizons of prompt neutrino ones. Tau air-showers  maybe born ( one over ten) also by resonant anti-neutrino electron at $6.4$ PeV energy. Therefore Tau Air-showers may be even born  without Tau neutrino at all.

\begin{figure}
\centering
\includegraphics[width=80mm]{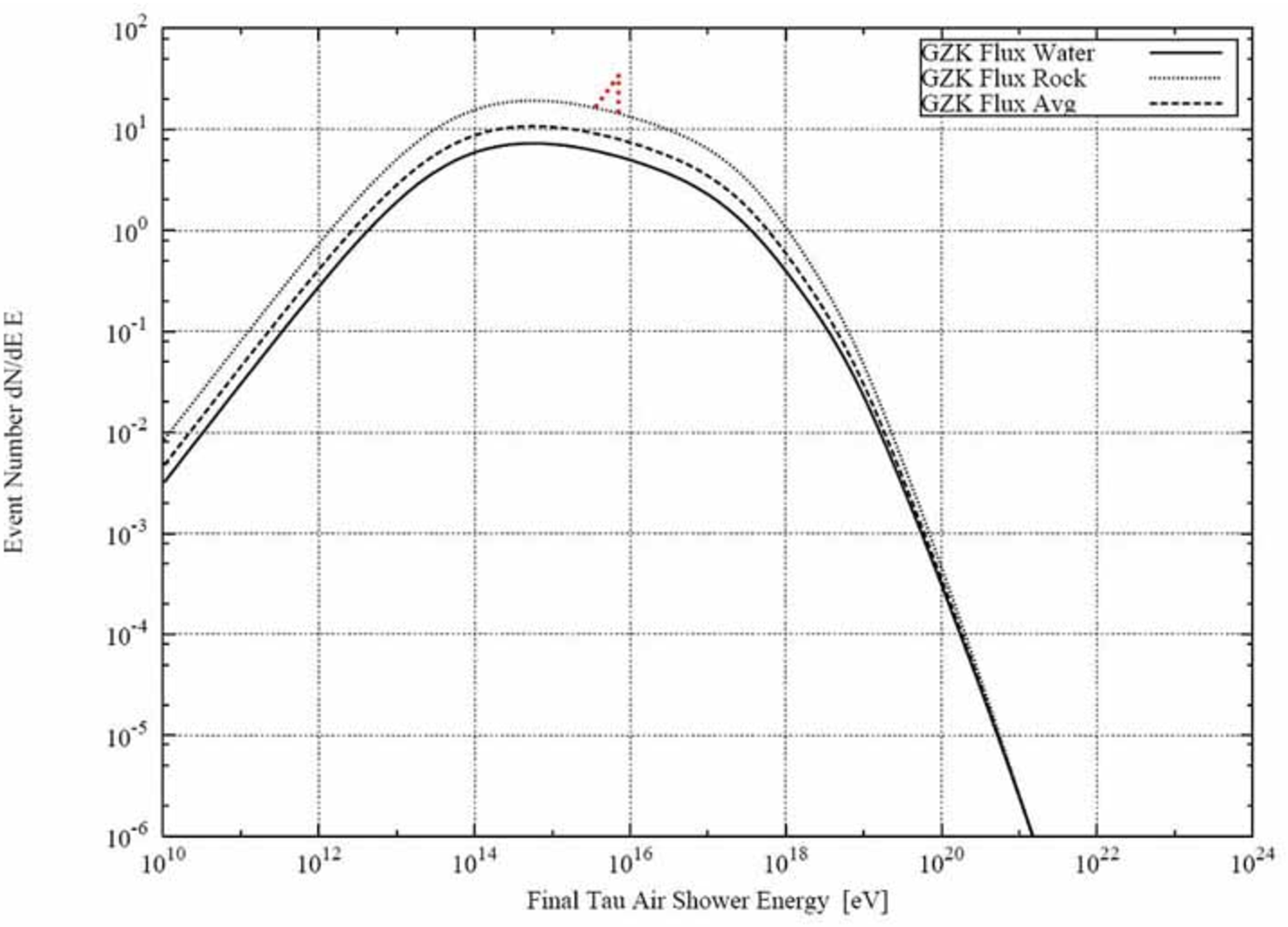}
\caption{ The integral rate of up going Tau including the resonant neutrino contribute in three years estimated in earlier papers. The small triangular bump due to Glashow resonant antineutrinos scattering on electrons is mainly due to the energy spread of the tau energy at its  birth. The rate are estimated for a total AUGER area at  10$\%$ efficiency for FD. The FD threshold is growing linearly with energy and it is respectively $300 Km^2$, $30 Km^2$, $3 Km^2$ at the energy $EeV$, $0.1 EeV$, $0.01$ EeV. Because of it each rate at lower energies than EeV must be suppressed in FD efficiency by corresponding factor (0.1, 0.01) making the expected event rate at ten PeV ($0.26$) and hundred PeV (0.5), just  below unity in 3 years. The additional mini-array AMIGA at $27.5 Km^2$, $6 Km^2$, whose array spacing is respectively $750$, $433$ m,  is a SD active day and night and it might double the signal, offering a detection in a very few   years from now  (0.35-0.4 event/year).} \label{Integral}
\end{figure}

  \section{ The UHECR GZK neutrino flux and  Tau expected rate}

  The predictions on Tau Astronomy moved fast from  sceptical  to a more  optimistic attitude
  from AUGER group, mostly via ground detectors (see figures below). On the contrary we  foresaw an event in FD within a few recent years
  since a few years with stable attitude. We have been optimistic from the earliest time foreseing one event within three years, if UHECR at WB flux was due to nucleons. However the very possible He composition of UHECR  reduce the EeV neutrino rate and correspondingly our predictions  by a factor two (at least). At lower energy there are signals at Tens PeV  observable in smaller detectors like HEAT and AMIGA and-or in FD telescope at nearer distances (2-3 km) to be discussed later.

\section{ Foreseen  Hadron Air-Showering via Fluorescence and Cherenkov }
 Tau Air-showers rise at horizons. Therefore it is important to test inclined downward hadronic air-showers. We did suggest
 first that the search of horizontal airshowers may be implemented by flashing cherenkov photons (rich signal but beamed one)
 and-or by fluorescence lights \cite{Fargion2007}. The prediction of such rare inclined events observable in both way has been offered just in earlier paper \cite{Fargion2006} and observed just a  year later by AUGER group \cite{Auger07}. The possibility to use both signals may offer a road map to calibrate the AUGER CR at lower energy. Indeed the Cherenkov flashes may rich PeV energies and they may at best pointing the Ande mountain chain. Therefore our hopes for a wild use of the AUGER and HEAT telescopes also for Cherenkov signal is  alive. The enhanced position of water-Cherenkov tanks nearby these telescopes  may help to test and increase their ability in inclined air-shower at horizons. Among them the upgoing Tau airshowers may still shine a little downward by the soft electromagnetic tails bent by geomagnetic fields. These possibilities maybe exploited at best toward the West (Ande) side.

\begin{figure}
\centering
\includegraphics[width=.45\textwidth]{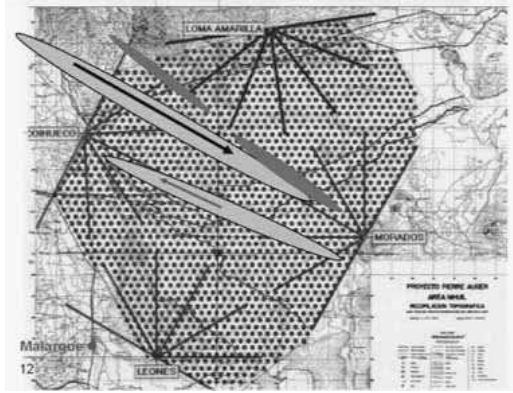}
\caption {Foreseen Inclined Horizontal Air-Showers able to trigger $both$
Auger tanks $and$ Fluorescence telescopes , while being in the
same axis. This technique, has never been
used to better disentangle horizontal Air-Showers. It may be an
ideal detector to observe Tau Air-Showers from the Ande. Their
events will populate the forbidden area of large zenith angle at
horizons. For this reason it will be useful to: a) enlarge the
angle of view of Coiuheco (as well as Loma Amarilla and Leones
station) toward the Ande; b) to eliminate any optical filter for Cherenkov lights in those directions ;
 c) to open a trigger between the Array-Telescope, or Telescope-Telescope in Cherenkov blazing mode; d) to try all $4$
telescope Fluorescence connection in Cherenkov common trigger-mode along all the $6\cdot 2 = 12$ common arrival directions.  Similar
connection along the $360^o$ view of stereoscopic HIRES telescopes, might be already recorded . Similar mutual use may rise in MAGIC-HESS-VERITAS array facing their telescopes at horizons triggered by vertical downward air-showers. In the picture some possible inclined UHECR
 events imagined (more than a year before the discovery) shining both array FD detectors and  (by Cherenkov lights) Fluorescence Station in AUGER;
  possible twin separated ovals arise by geomagnetic bending has been partially observed.}\label{fig:fig1}
  \end{figure}
   \vspace{9pt}
 \begin{figure}
 \centering 
  \includegraphics[width=.45\textwidth]{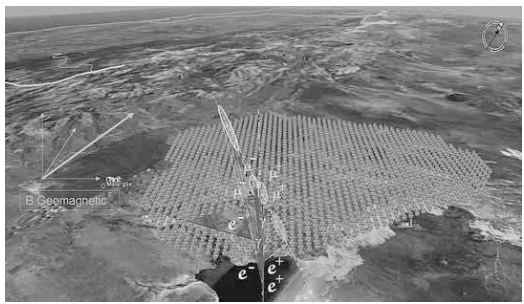}
\caption {As above the observed  inclined air-shower as expected above to hit both at
Cherenkov and Fluorescence track an year later (2007) by the AUGER group . More details in (Fargion at all. 2007-NIM)}\label{fig:fig1}
\end{figure}
\newpage

\section{ UHECR  Lightest Nuclei versus UHE $\nu$: the GZK connection}
Tau decay in flight are better than neutrino interacting in air  because Earth rock density is three thousand times larger than air one. Even if the tau enjoy of a bounded escaping solid angle (a zenith angle width of $2-5^°$ degree) while all down-ward neutrinos interacting on air may reach from wider cone (a zenith angle width  $15^°$ degree) and are in three flavors. Such skimming events in AUGER experiment may rise via upward tau air-shower \cite{Fargion1999},\cite{Bertou2002},\cite{Bigas07}. In last years the upward-horizontal EeV $\tau$,$\bar{\tau}$ appearance, via UHECR $p + \gamma_{CMB}\rightarrow \pi \rightarrow \nu $  has been predicted by many authors; the most extreme ones were at rate of $0.1-0.03$ a year \cite{Bigas07}, or  $0.3$ a year in AUGER \cite{Fargion2007} and finally up to $0.2$ a year \cite{Auger08}. This rate, has only recently being adjusted and confirmed by last AUGER group estimates (NOW 2008, CRIS 2008):$0.3$ a year, in full agreement with our previous (and persistent) ones \cite{Fargion2007} . The last AUGER predictions are even overcoming our expected rate \cite{Auger09} even in over optimistic way. Indeed following the   AUGER evidences (and Hires ones \cite{Hires06})  of  an UHECR GZK cut-off and the latest AUGER (possible) Super-Galactic anisotropy due to an eventual proton UHECR guarantee a secondary flux of UHE-GZK neutrino at EeVs energy within AUGER detection via $\tau$,$\bar{\tau}$ showering \cite{Fargion2007}. In fact this occurs because the muon neutrino flavor mixing must feed also a  tau neutrino component. Such UHE astrophysical tau neutrino (noise-free  from any atmospheric background) may interact in and it may rises out the Earth as UHE $\tau$. The UHE $\tau$,$\bar{\tau}$ decay in flight in atmosphere must lead to loud Tau Air-showers. Such a detectable flashes may rise in short times within Auger SD (by large electromagnetic curvature signals) or in FD arrays by horizontal fluorescence signals, namely once in a few years ($2-4$) from now \cite{Fargion2007}, \emph{if the UHECR are nucleons}.

 \section{ UHECR from Cen-A and the Lightest Nuclei spreads}
 Nevertheless a recent alternative UHECR understanding \cite{Fargion2008}, based on observed AUGER UHECR  (nuclei) mass composition and with Cen-A rich clustering map, is in disagreement with UHECR proton understanding \cite{Auger-Nov07}. This UHECR understanding is leading to different UHE neutrino  predictions. It suggests that UHECR are made by Lightest Nuclei ($He^2_{4},He^2_{3} $, maybe also $Li$,$Be$), mostly originated from Cen-A.

  \begin{figure}
\centering
\includegraphics[width=.5\textwidth]{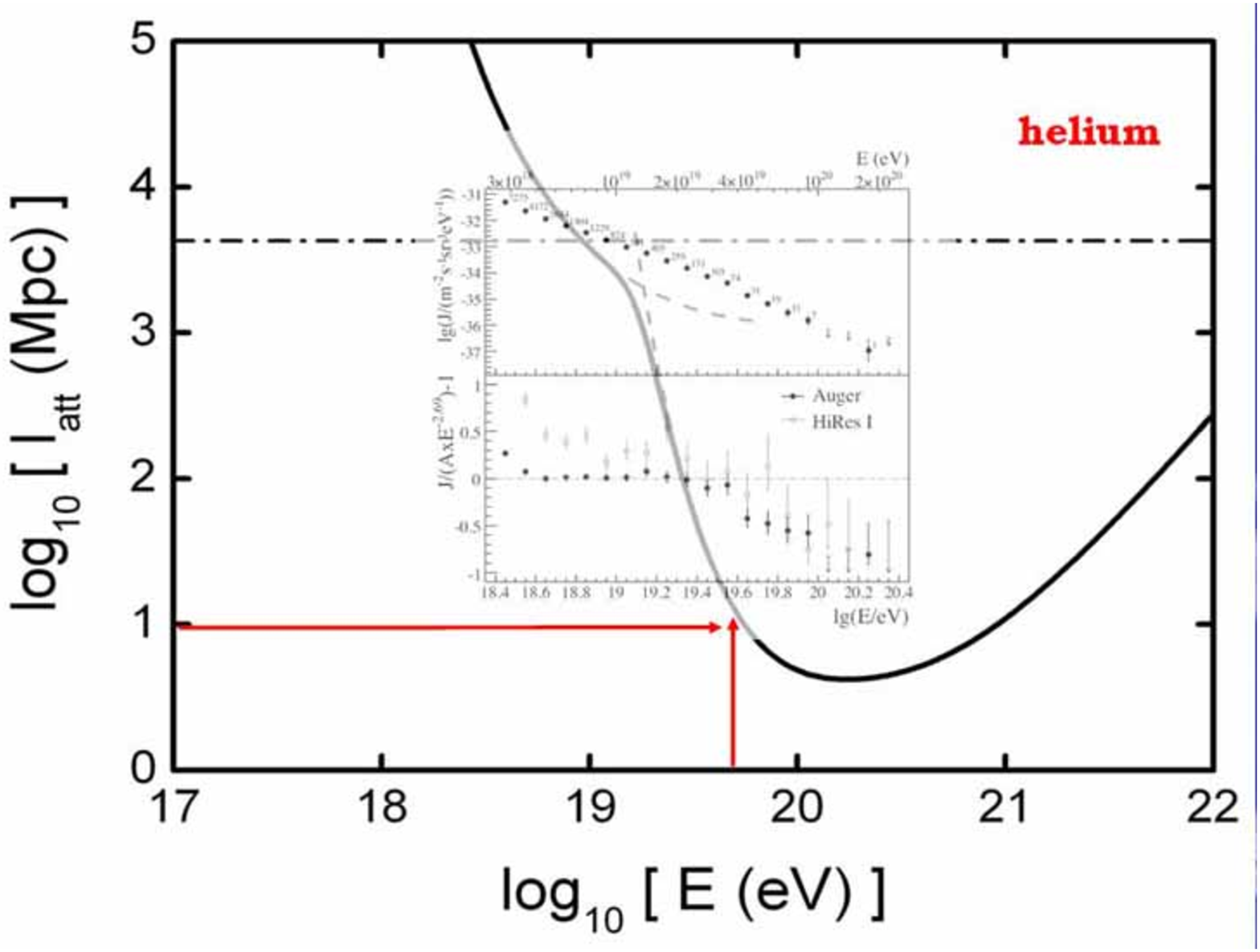}
\caption{The Helium  free length is more bounded than proton or iron nuclei. Its sharp decrease may better fit the observed GZK knee. Its short distance may explain the Virgo (up to day) absence. The UHECR arrival spread for any given energy will be proportional to the Z charge of Lightest nuclei. For a given energy the UHECR spread in groups around Cen-A  may offer a first spectroscopy of lightest nuclei UHECR composition}See
ref.\cite{Fargion2001}\label{fig:HE-GZK}
\end{figure}

  The HE-like UHECR trajectories are bent and spread by spiral horizontal galactic magnetic fields into vertical spread axis. Therefore UHECR appear as  (incidentally) clustered  ( by galactic fields) around Cen-A just along the Super-Galactic Arm. Cen-A is possibly an unique nearest source able to survive the short Lightest Nuclei GZK cut-off. Virgo is too far (and a little out of view of AUGER detector). These events spread mostly along the same  Super-Galactic Arm  apparently from far $80$ Mpc  Centaurs Cluster. The mean random angle  bending $He^2_{4}, Li^3_{6},Be^4_{8} $,  by spiral galactic magnetic  fields along the plane are easily found  $\delta_{rm} \geq$:
\begin{equation}
{11.3^\circ}\cdot \frac{Z}{Z_{He^2}} \cdot (\frac{6\cdot10^{19}eV}{E_{CR}})(\frac{B}{3\cdot \mu G})\sqrt{\frac{L}{20 kpc}}
\sqrt{\frac{l_c}{kpc}}
\end{equation}

\begin{equation}
{16.95^\circ}\cdot \frac{Z}{Z_{Li^3}} \cdot (\frac{6\cdot10^{19}eV}{E_{CR}})(\frac{B}{3\cdot \mu G})\sqrt{\frac{L}{20 kpc}}
\sqrt{\frac{l_c}{kpc}}
\end{equation}

\begin{equation}
{22.6^\circ}\cdot \frac{Z}{Z_{Be^4}} \cdot (\frac{6\cdot10^{19}eV}{E_{CR}})(\frac{B}{3\cdot \mu G})\sqrt{\frac{L}{20 kpc}}
\sqrt{\frac{l_c}{kpc}}
\end{equation}

\begin{figure}
\centering
\includegraphics[width=.7\textwidth]{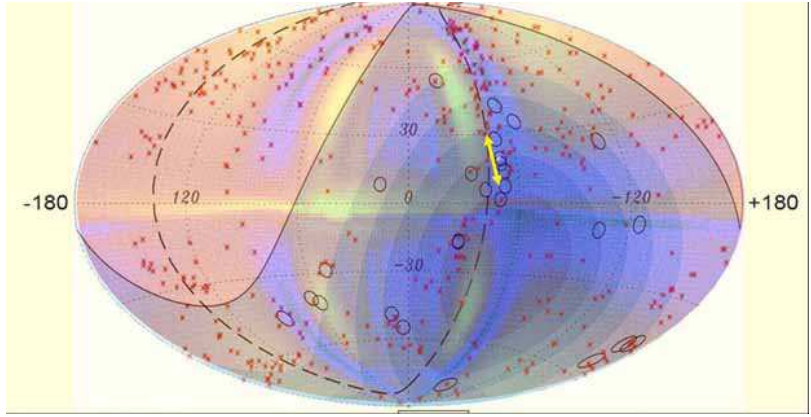}
\caption{The first order bending for lightest UHECR, as He nuclei, is shown by the vertical arrow. The underline
galactic magnetic fields are spiral lines on Galactic Plane spreading, by Lorentz force, vertically in Cen-A region,  explaining the longitudinal
clustering of the events, overlapping by chance on Super Galactic Arm , shown by the dashed curve. It also explain the UHECR composition and the possibly absence of Virgo.  For a given energy (as the chosen threshold one)  the UHECR spread in groups around Cen-A  may offer a first spectroscopy of few  UHECR lightest nuclei}\label{fig:fig1}
\end{figure}

This {\emph{Lightest Nuclei for Highest Cosmic Rays}} model  implies and foresees among the other, additional clustering of UHECR events around the nearest AGN Cen-A (the lightest UHECR the more correlated to the source, the heavier and with larger charges, the most bent and spread ones). The model        explains the absence or a poor  signal  from Virgo (too far for the fragile nuclei to fly by) and possibly also from Fornax. Such {\emph{Lightest Nuclei for Highest Cosmic Rays}}  are forced in a very confined cosmic volume (ten Mpc or less) due to a fragile light nuclear (few MeVs) binding energy. Usually heavy nuclei fragmentation pour energy only  in UHE neutrino (at $0.1-0.01$ EeV energy) spectra, less energetic than common UHE EeV $p+\gamma_{CMB}\rightarrow\pi$ neutrino flux.

\section{UHECR by Lightest Nuclei versus $\nu_{\tau}$ }
AUGER SD or FD are not able to reveal tens-hundreds PeV energy  easily. However a very recent and a less spaced AUGER sub-system, a more dense array AMIGA and the additional telescopes HEATS might lower the threshold accordingly. The $0.1 EeV$ mostly hadronic, inclined-\emph{upward}, tau air-showers $\theta \leq 80°$ occur at much lower altitudes than hadronic inclined \emph{down}-going air-showers,
at much nearer distances from telescopes than hadronic EeV air-showers, reducing the area and the rate. They may offer a tens-hundred PeV neutrino windows, secondaries of UHE He nuclear fragmentation. Some estimates are  offered assuming, for sake of simplicity a Fermi-like
UHE GZK spectra, comparable to the well known Waxmann-Bachall flux derived for cosmic GRBs.  Nevertheless if rarest UHECR, possibly from AGN,  above $1-2 \cdot 10^{20}  eV$, ( a few in AUGER events), are also He nuclei, they may still suffer GZK photo-pions opacity decaying into EeV neutrinos too. Such He GZK interaction  may be comparable with corresponding proton GZK ones at $6 \cdot 10^{19} eV$: therefore EeV tau Neutrino Astronomy, tails of the most energetic Lightest Nuclei GZK secondaries, may still rise at AUGER.  EeV inclined tau decays, born nearly at sea level (versus inclined hadronic $\theta \geq 80°$ ones developing at altitudes well above ten km) occur at air-density  at least three times higher (than hadron horizontal ones); therefore  Tau Air-showers (mostly behaving as hadron ones), their electromagnetic and fluorescence tail  sizes (and times) are three times shorter ($l_{sh}\approx8-10$ km) and their characteristic azimuth speed  $\dot{\varphi}\approx 1.6 \cdot 10^4 rad s^{-1}$ is slower than common inclined UHECR hadron ones at higher altitudes ($l_{sh}\geq 25-30$ km), ($25-35$ km),$\dot{\varphi}\geq 5 \cdot 10^4 rad s^{-1}$. Such a brief $20-30\mu$s and distinct  up-ward tau lightening versus slow $60-90 \mu$s diluted down-ward horizontal showering  may be  well disentangled within AUGER (and AMIGA SD) threshold, also by obvious angular resolution and directionality (up or downward); their strong curvature and their inclinations would rise within HEAT FD and AUGER FD, we estimate, once every a (2-3)  years. Possibly (factor two) originated from the  Ande side (West to East )\cite{Fargion1999},\cite{aramo05},\cite{Miele06}.

 \begin{figure}[htb] \vspace{9pt}
 \centering
\includegraphics[width=70mm]{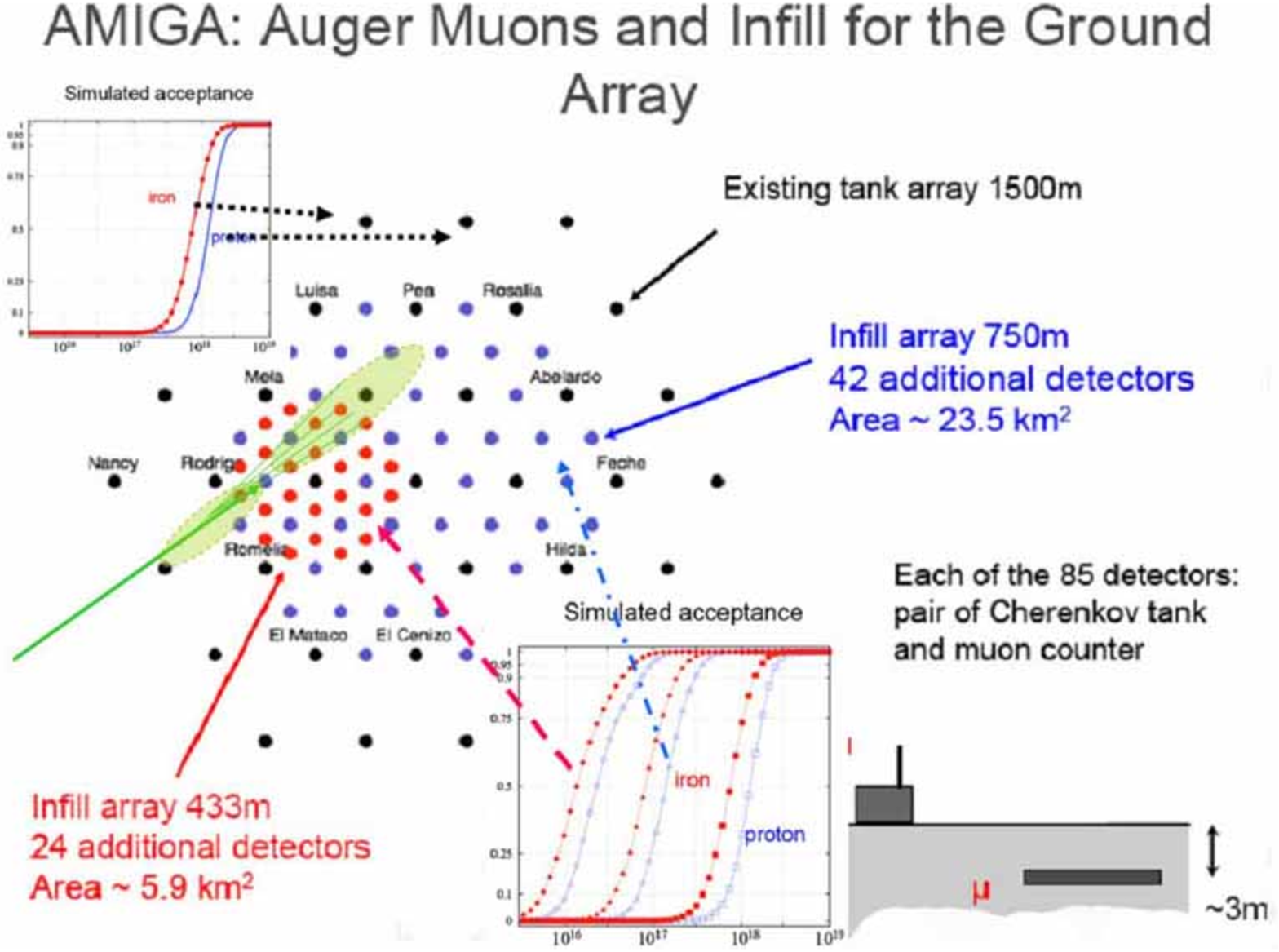}
\caption{ The AUGER Array and the inner area AMIGA where lower energy air-showers might be revealed by ground detectors.
The two different spacing (5.9 $km^2$ and the 23.5 $km^2$) offer a lower energy threshold for neutrino inclined air-showers respectively at $10^{16}$ and $10^{17}$ eV . In the figure a schematic area due to air-showering lobes of an escaping horizontal air-shower } \label{Array}
\includegraphics[width=70mm]{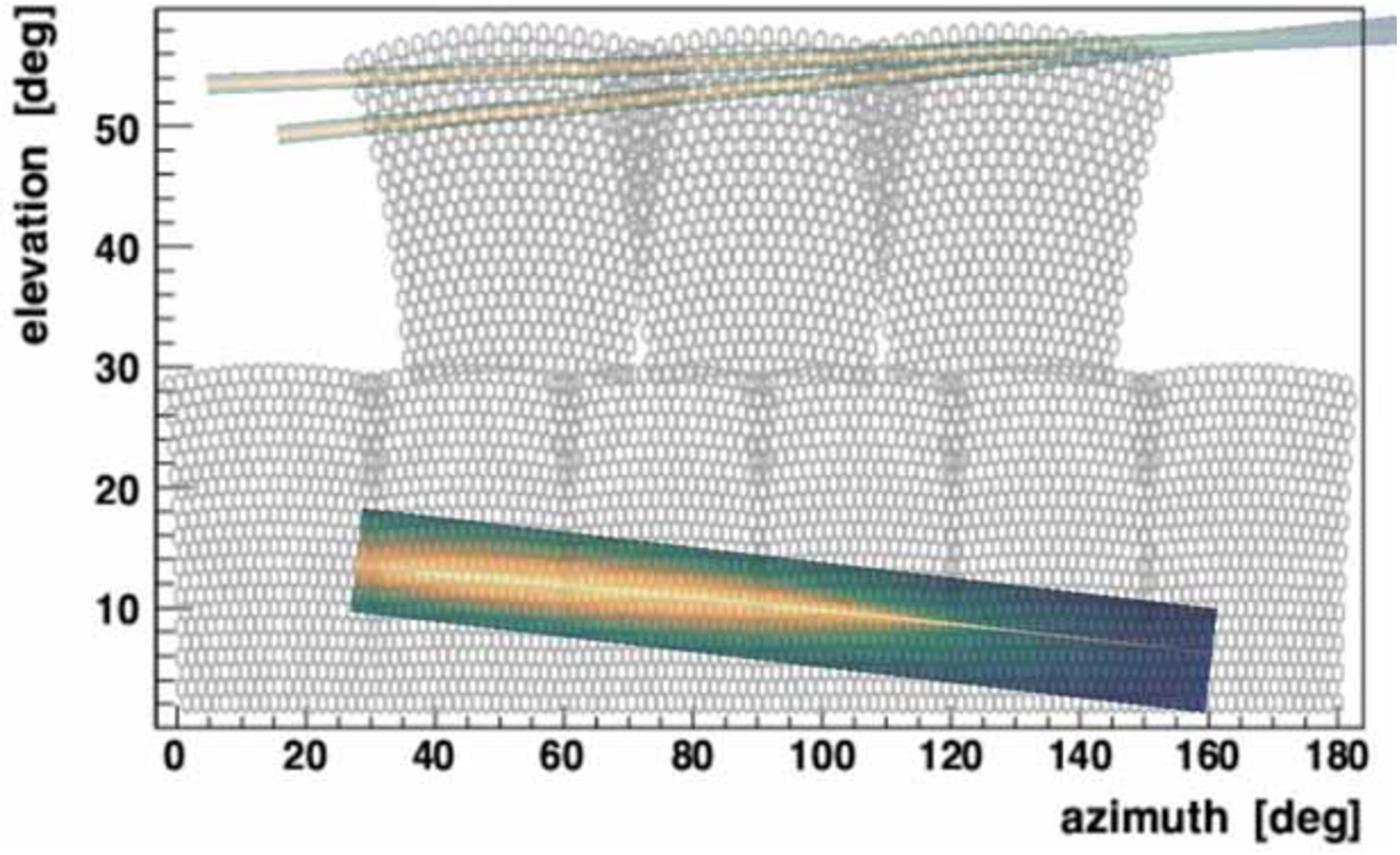}
\caption{ As above the inner HEAT where lower energy air-showers might be revealed by Fluorescence telescopes.
 In the figure a schematic array phototube where air-showering lobes of a near and escaping Tau horizontal air-shower at few ten PeV energy. Tau will be raised at few km (2-3) if at PeVs energy and at tens km (20-30) if of EeV nature.  A more common downward-horizontal hadronic airshower at tens EeV may split by geomagnetic fields shining at far distances and very high altitudes (25-35) km. Almost ever outside the AUGER ground detector area. The altitude, timing, angle spread are unique imprint. They may serve to disentangle Nuclei-versus nucleon UHECR nature.} \label{Array}
\end{figure}
This new neutrino Astronomy, the PeVs-EeVs ratio, may disentangle in future records  the real UHECR  nucleon or  lightest nuclei UHECR nature. Indeed the additional rise  of the resonant Glashow  contribute, $\bar{\nu}_e+ e \rightarrow W^-\rightarrow \bar{\nu}_{\tau} + \tau $, in the upgoing Tau, while just marginally doubling the signal at 0.01 EeV. Leading to a flavor UHE neutrino spectroscopy.
 \subsection{The resonant $\bar{\nu}_e+ e \rightarrow W^-\rightarrow \bar{\nu}_{\tau} + \tau $}
   The role of UHE neutrino interaction with matter is well understood: they also shape the neutrino
   survival across the Earth. Indeed the highest energy $\nu$ are opaque to Earth, but not to
   smallest cord. Therefore the harder events are the more tangent ones. The
    resonant $\bar{\nu_e} + e \rightarrow W^-\rightarrow \bar{\nu_{\tau} + \tau}$ are very peculiar
    signals. Their opacity on Earth is extreme. They are as opaque as EeV energetic neutrinos.
    But their lower energy corresponds to higher flux (for constant energy fluency as Waxmann Bachall one).
    But their propagation in Earth is much smaller too. The compensated flux in 5-6 times higher than EeV one.
    Unfortunately this rate is (in AUGER) suppressed by detection threshold (nearly one hundred times smaller
    than EeV showers). Nevertheless the mini array AMIGA (of 6 $km^2$) and the HEAT telescopes may contribute
    to make detecable in a few year the same resonant  $\bar{\nu_e} + e \rightarrow W^-\rightarrow \bar{\nu_{\tau} + \tau}$ :
    the distance will be ten times smaller and the  showering azimuth angular velocity will appear $ten$ times faster
    than already fast EeV Tau Airshower. Their curvature in AMIGA may leave a clear imprint.

  \begin{figure}
\centering
  \includegraphics[width=.4\textwidth]{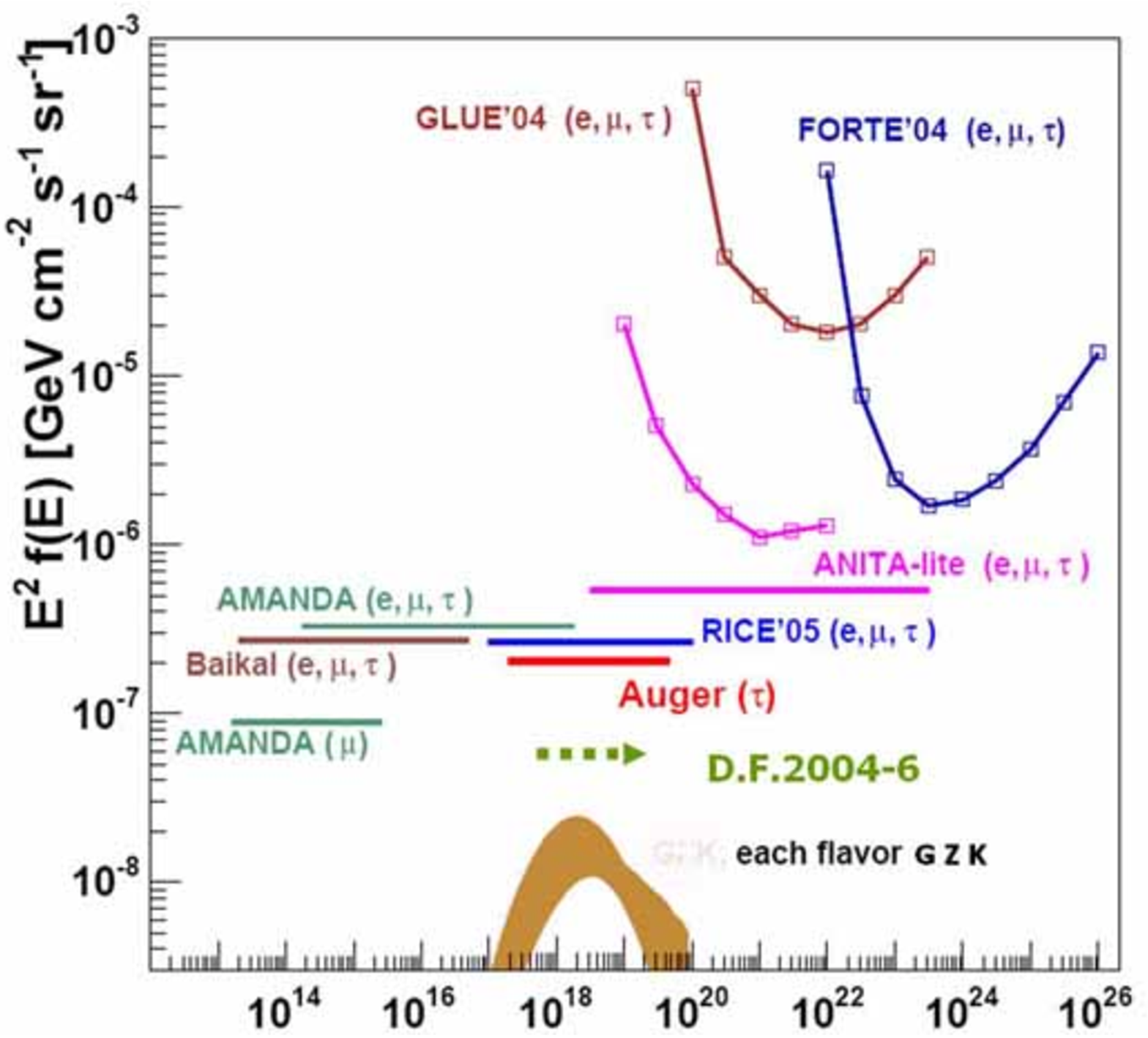}
\caption {The earliest prediction of AUGER group versus expected fluxes (ICRC-2005) by AUGER group. Note a nominal 10-20 years
duration for an event Tau versus our 3-4 years rate.}\label{fig:fig1}
  \includegraphics[width=.4\textwidth]{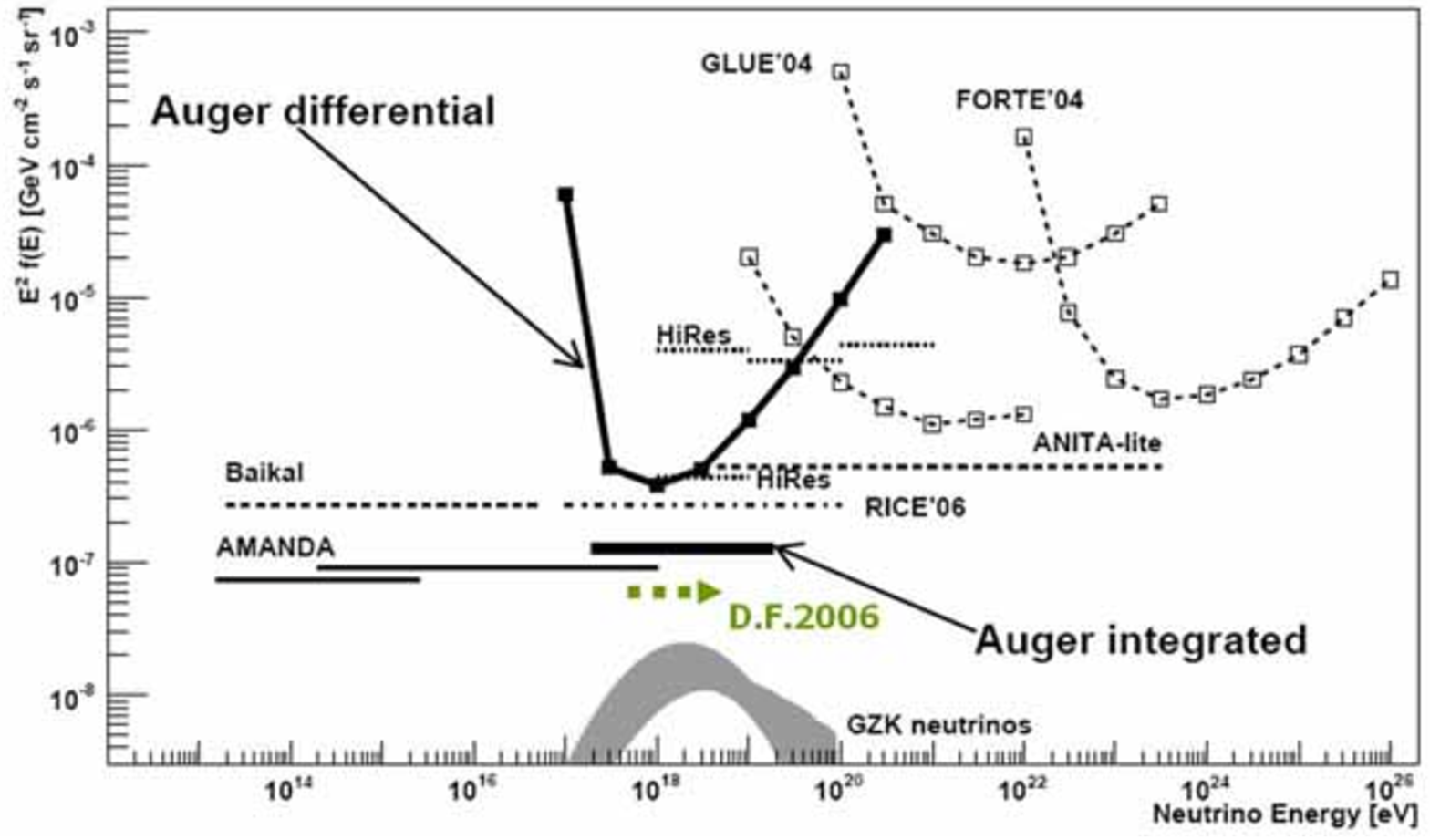}
\caption {The recent prediction of AUGER group versus our earlier expected fluxes (ICRC-2007) by AUGER group. Note a nominal 6 years
duration for an event Tau}\label{fig:fig1}
 \includegraphics[width=.5\textwidth]{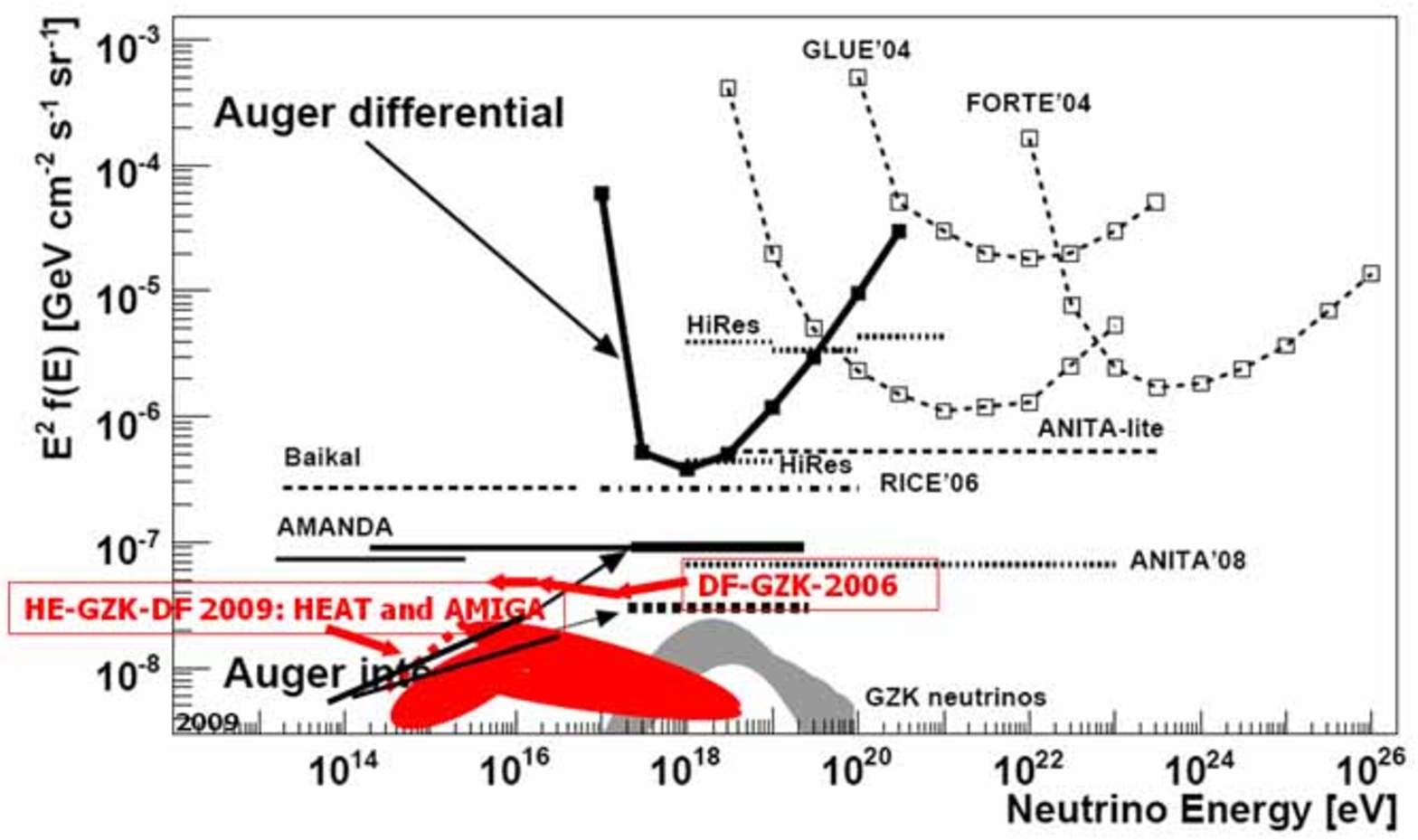}
\caption {The latest prediction of AUGER group (2009), versus expected fluxes by our past and present paper, based also on possible UHECR HE-GZK nature. Note a nominal $1-4$ years rate for an event, if UHECR are nucleon, versus a three years at tens PeV, for a UHECR-He nuclei}\label{fig:fig1}
\end{figure}

\section{Conclusions: UHE $\nu_{\tau}$ Astronomy by UHECR beyond the corner}
UHECR Astronomy is a new windows achieved last few years. We believe that UHECR  signal rose with definitive direction  mostly from Cen-A. We explained why He-like lightest nuclei solve most composition puzzles and the smearing nature of these CEN-A events. We foresee (as soon data by AUGER will be released): a) additional UHECR clustering along Cen-A in the "vertical" axis (as earlier events), testing somehow the light nuclei composition. b) the possible emergence of a smeared Virgo if and only if also heavier  $C-N-O$ nuclei are ejected; otherwise He-like are the main currier; c) the possible more smeared presence of diffused isotropic events as fragments of far sources (with little astronomical meaning).d) The Virgo might be soon
a hidden (by the Earth shadows) source of skimming UHE HE-GZK neutrinos, at best beyond the Ande. Either by \emph{EeV GZK \cite{Greisen:1966jv},\cite{za66} secondaries if AUGER did reveal nucleons } or, following light nuclei model,  by tens PeV signals. Their detection via FD in AUGER and HEAT is, we believe, at the edge. The short and near air-shower from the FD telescope (due to the lower energy thresholds) makes their Tau airshower timing brief, sharp and at small zenith angle,as well as up-going.
On the contrary the rarest inclined horizontal down-going hadron shower are far away, diluted in air density
and spread in much longer time scale. They are  often  bifurcate (by geomagnetic field). No way to be confused.
Three or more times duration and morphology of \emph{up or down} signature will disentangle any rare neutrino lights arriving from their
unique but unusual sky: the Earth. Their rate, timing, energy and inclination may teach us on the real nature (nucleon or lightest nuclei) of
UHECR. In this hope we suggest (a) to implement the present AUGER array with additional array searching for shadows of inclined hadronic from Ande in AUGER; (b) To add Mini-Cherenkov telescope arrays facing Ande screen to reveal tau air-showers signals beyond the mountain chain also via direct Cherenkov flashes.(c) Locating a new telescope array few km distance to reduce the energy thresholds (d) Introducing novel trigger detection by fast horizontal track time signature.(e) Locating SD nearest to FD in order to enlarge their ability to discover the inclined events  (blazing Cherenkov flashes) both in FD and in Cherenkov lights as well as in muon tracks in nearby SD array elements. It must be remind that the AMIGA muon versus
SD Cherenkov signal may well signal the electromagnetic versus muon nature of horizontal air-shower, offering an additional tool to disentangle neutrino versus  hadronic air-showers. We foresee in very near years, possibly this and next one, the Tau Neutrino Astronomy birth and blow-up, if accurate attention on trigger thresholds, and AMIGA-HEATS implementation will be concluded soon. Neutrino Astronomy by Tau exploit a natural amplifier, the air shower , as a telescope. Its direction and view is the deeper microscope in star or black-hole secret interiors.   Incidentally the birth of tau neutrino astronomy may be at the same time the first definitive probe of  neutrino  flavor mixing and reappearance.  \\



\end{document}